\documentclass[structabstract]{aa}  
\usepackage{graphicx}
\usepackage{longtable,lscape}
\usepackage{natbib}
\usepackage{txfonts}

\def\asec{\ifmmode ^{\prime\prime}\else$^{\prime\prime}$\fi}
\def\farcs{\hbox{$.\!\!^{\prime\prime}$}}       
\def\degs{\ifmmode ^{\circ}\else$^{\circ}$\fi}

\begin{document}

\titlerunning{BL~Lac objects beyond redshift 1.3 - UV-to-NIR photometry and photometric redshift for Fermi/LAT blazars}
\title{BL Lacertae objects beyond redshift 1.3 - UV-to-NIR photometry and photometric redshift for Fermi/LAT blazars}

   \author{A. Rau,
          \inst{1}
          \and
          P. Schady,
          \inst{1}
          \and
          J. Greiner
         \inst{1}
          \and
          M. Salvato\inst{2,3}
          \and
          M. Ajello\inst{4,5}
          \and
          E. Bottacini\inst{4}
          \and
          N. Gehrels\inst{6}
          \and 
          P.~M.~J. Afonso\inst{1,7}
          \and
         J. Elliott\inst{1}
          \and
          R. Filgas\inst{1}
          \and
          D.~A. Kann\inst{8}
          \and
          S. Klose\inst{8}
          \and
          T. Kr\"uhler\inst{1,3,9}
          \and
          M. Nardini\inst{1,10}
          \and
          A. Nicuesa Guelbenzu\inst{8}
          \and
          F. Olivares E.\inst{1}
          \and
          A. Rossi\inst{8}
          \and
          V. Sudilovsky\inst{1}
          \and
          A.~C. Updike\inst{11}
          \and
         D.~H. Hartmann\inst{12}
         }

   \institute{Max-Planck-Institut f\"ur Extraterrestrische Physik, Giessenbachstra{\ss}e 1, 85748 Garching, Germany\\
              \email{arau@mpe.mpg.de}
         \and
      Max-Planck-Institut f\"ur Plasma Physik and Excellence
     Cluster, Boltzmannstrasse 2, 85748 Garching, Germany
     \and
     Universe	Cluster,	Technische	Universit\"at	M\"unchen,	Boltzmannstra{\ss}e	2,	85748	Garching,
     Germany
      \and
     W.W. Hansen Experimental Physics Laboratory \& Kavli Institute for Particle Astrophysics and Cosmology, Stanford University, USA
        \and
     SLAC National Accelerator Laboratory, StanfordUniversity, Stanford, CA 94305, USA 
         \and
     NASA-Goddard Space Flight Center, Greenbelt, Maryland 20771, USA
     \and
     American River College, Physics \& Astronomy Dpt., 4700 College
     Oak Drive, Sacramento, CA 95841
    \and 
     Th\"uringer Landessternwarte Tautenburg, Sternwarte 5, 07778
     Tautenburg, Germany
    \and
     Dark Cosmology Centre, Niels Bohr Institute, University of
     Copenhagen, Juliane Maries Vej 30, 2100 Copenhagen, Denmark
         \and
     Universit\'a degli studi di Milano-Bicocca, Piazza della Scienza
     3, 20126, Milano, Italy
     \and
     Department of Physics and Astronomy Dickinson College Carlisle,
     PA 17013, USA 
     \and
    Department of Physics and Astronomy, Clemson University, Clemson, SC 29634, USA
            }

   \date{Received September 27, 2011; accepted November 19, 2011}

 
  \abstract
   {Observations of the $\gamma$-ray sky with {\textit Fermi} led to
     significant advances  towards understanding blazars, the most extreme class
     of Active Galactic Nuclei. A large fraction of the population
     detected by {\textit Fermi} is formed by BL~Lacertae  (BL~Lac)  objects, whose
     sample has always suffered from a severe redshift incompleteness
     due to the quasi-featureless optical spectra.}
   {Our goal is to provide a significant increase of the number of
     confirmed high-redshift BL~Lac objects contained in the 2~LAC
     {\textit Fermi}/LAT cataloge.}
   {For 103 {\textit Fermi}/LAT blazars, photometric redshifts using spectral energy
     distribution fitting have been obtained. The photometry includes
     13 broad-band filters from the far ultraviolet to the near-IR
     observed with {\it Swift}/UVOT and the multi-channel imager GROND
     at the MPG/ESO 2.2m telescope. Data have been taken
     quasi-simultaneously and the remaining source-intrinsic
     variability has been corrected for.}
   {We release the UV-to-near-IR 13-band photometry for all 103
     sources and provide redshift constraints for 75 sources without
     previously known redshift. Out of those, eight have reliable
     photometric redshifts at $z\gtrsim1.3$, while for the other 67
     sources we provide upper limits. Six of the former eight are
     BL~Lac objects, which
     quadruples the sample of confirmed high-redshift BL~Lac. This
     includes three sources with redshifts higher than the previous
     record for BL~Lac, including CRATES~J0402-2615, with the best-fit
     solution at $z\approx1.9$.}
   {}


   \maketitle
%


\section{Introduction}

Since  its  launch in  2008,  the  {\it  Fermi} Space  Laboratory  has
dramatically extended  our view of the high-energy  sky.  The recently
released  24-month catalog  (2LAC)  of Active  Galactic Nuclei  (AGN)
detected  by the Large  Area Telescope  \citep[LAT; ][]{Atwood:2009aa}
revealed  885 high-significance  sources, the  large majority  of them
being  blazars  \citep{Ackermann:2011lr}.  The  latter  form the  most
extreme class of AGN with their observational characteristics governed
by the small angle between  their relativistic jets and the observer's
sight line  \citep{Blandford:1978lr}.  The resulting  Doppler boosting
makes blazars  exceptionally bright sources at  nearly all wavelengths
and therefore visible out to high redshift.

The  scientific  relevance of  blazars  is  very  broad, ranging  from
laboratories for  the physics and structure of  relativistic jets (and
thus the  extraction of  energy from the  central massive  black hole)
\citep[e.g.,][]{Abdo:2010ab}   to   probes   of   the   extra-galactic
background  light (EBL)  through attenuation  of  $\gamma$-ray photons
\citep[e.g.,][]{Abdo:2010ac}.  One of the crucial parameters for these
applications is  the distance to  the objects, which  unfortunately is
not easy to obtain in most cases.

Two classes  of blazars  dominate the 2LAC  population. These  are the
flat-spectrum radio  quasars (FSRQs,  310 sources) and  BL~Lac objects
(395),      named       after      the      prototype      BL~Lacertae
\citep[][]{Hoffmeister:1929lr}.    While  for   the   former  redshift
measurements are routinely performed using their strong emission lines
at UV-optical wavelengths,  the featureless, power-law optical spectra
of    BL~Lac    objects    have    proven   to    be    a    challenge
\citep[e.g.,][]{Shaw:2009fk}. Indeed,  220 of  the 395 BL~Lacs  in the
2LAC (55\,\%)  lack redshift estimates.  Until  this incompleteness is
resolved,   conclusions   about   the   EBL,   the   blazar   sequence
\citep[e.g.,][]{Fossati:1998lr,Ghisellini:1998fk},   and   the  blazar
population in general remain tentative at best.

Several methods  have been exploited  to increase the  BL~Lac redshift
sample.   At low  distance,  one can  utilize  the remarkably  uniform
absolute  brightness  of the  giant  elliptical  BL~Lac host  galaxies
\citep{Sbarufatti:2005qy,Meisner:2010uq} and very-high signal-to-noise
optical spectroscopy to help  to identify weak emission or absorption
features  in  a  few  other  cases  \citep[e.g.,][]{Shaw:2009fk}.   An
alternative  method, applicable  to the  more distant  sources,  is the
photometric  redshift technique, which consists of fitting spectral
energy distribution (SED) templates to multi-band photometry.  Neutral hydrogen  along the  line of
sight to the blazar will  imprint a clear attenuation signature at the
Lyman limit and thus allows an accurate estimate of the redshift of the
absorber.  Even  though the absorber  will be located  somewhere along
the line of sight and its redshift will thus not necessarily correpond
to that of  the blazar, photometric redshifts, $z_{\rm phot}$, will  provide a reliable
lower limit on the blazar redshift. 

In  this  paper  we  explore  the  use  of  ultra-violet  to  near-infrared
quasi-simultaneous  photometry to  measure the  redshift of  103 2LAC
blazars, 86 of them without previous redshift constraints.  The sample
selection is  described in Sect.~\ref{sec:sample}  while the observations
and   data   reduction    are   detailed   in   Sect.~\ref{sec:obs}   and
Sect.~\ref{sec:data}. The  results and their discussion  are presented in
Sect.~\ref{sec:results} and Sect.~\ref{sec:discussion}, respectively.


\section{Sample Selection}
\label{sec:sample}

The sample is selected mainly  from the clean sample of the 24-month
catalog  of AGN detected  by {\it  Fermi}/LAT \citep{Ackermann:2011lr}
which  includes 885 high-significance  $\gamma$-ray sources  which are
statistically  associated  with  AGNs  and located  at  high  Galactic
latitudes ($|b|>10$\,deg).   To accomodate the location  of one of
our follow-up instruments in the southern hemisphere, we applied a cut
in  declination  ($\delta_{\rm  J2000}< +25$\,deg).   Furthermore,  the
importance of the ultra-violet  bands for the $z_{\rm phot}$ estimates
lead us to discard all  sources with Galactic foreground reddening of
E$_{B-V}>0.2$\,mag, as  derived from the  \cite{Schlegel:1998ul} maps.
The final sample is composed  of 80 sources without known redshift but
with   optical  or  radio   counterpart  associations   from  Bayesian
statistics \citep{Abdo:2010lr,Abdo:2010fk}, Likelihood Ratio, or log N
-- log S methods  \citep{Ackermann:2011lr} and was further extended by
eight sources that do not belong to the clean 2LAC sample\footnote{flagged due
  to analysis issues  by the LAT team}. In  addition, observations for
16 2LAC sources with  existing redshift measurements are included for
verification.  We emphasize here that our sample is not statistically
complete, but biased by selection. The final  list  of  104  targets  is presented  in
Table~\ref{tab:sample}. Of  those, 82  have been classified  as BL~Lac
objects, three  as a FSRQ,  and the remaining  19 are of  unknown type
\citep[][]{Ackermann:2011lr}. 
                        
\begin{table*}
\caption{\label{tab:sample} Target List (first five rows, rest in Journal only)}
\begin{tabular}{llcccc}
\hline\hline
Name  & FGL Name$^a$ & $\alpha_{\rm J2000}$$^b$ & $\delta_{\rm J2000}$$^b$ & offset [$^{\prime\prime}$]$^c$ & E$_{B-V}$ [mag]$^d$\\
\hline
CRATES~J0001-0746 & 2FGLJ0000.9-0748 & 00:01:18.00 & -07:46:26.9 & 0.2 & 0.03 \\
CRATES~J0009+0628 & 2FGLJ0009.0+0632 & 00:09:03.93 & 06:28:21.3 & 1.1 & 0.07 \\
CRATES~J0021-2550 & 2FGLJ0021.6-2551 & 00:21:32.56 & -25:50:49.1 & 0.8 & 0.02 \\
1RXS~J002209.2-185333 & 2FGLJ0022.2-1853 & 00:22:09.26 & -18:53:34.9 & 3.1 & 0.03 \\
RX~J0035.2+1515 & 2FGLJ0035.2+1515 & 00:35:14.71 & 15:15:04.2 & 3.4 &
0.07 \\
.. & .. & .. & .. & .. & ..\\
\hline
\end{tabular}
\\
$^a$: {\it Fermi}/LAT ID.\\
$^b$: GROND coordinates of the optical counterpart with typical uncertainties of 0\farcs3 in both directions.\\
$^c$: Offset of counterpart location from those given in \cite{Ackermann:2011lr}. \\
$^d$: From \cite{Schlegel:1998ul}.\\
\end{table*}


\section{Observations}
\label{sec:obs}

The  observations  were performed  with  the  Ultraviolet and  Optical
Telescope  \citep[UVOT;][]{Roming:2005lr}   onboard  the  {\it  Swift}
satellite      \citep{Gehrels:2004fk}      and      the      Gamma-ray
Optical/Near-Infrared     Detector    \citep[GROND;][]{Greiner:2008aa}
mounted  on the  MPI/ESO 2.2\,m  telescope  at La  Silla, Chile.  Each
source has been observed at least once with both instruments, although
for some targets  several epochs had been acquired.  

The UVOT pointings
have been  obtained as  part of the  {\it Swift} fill-in
program from
January  2010  to  October  2011.  With  few  exceptions,  they  comprise
observations in three  optical ($u, b, v$) and  three UV ($uvw2, uvm2,
uvw1$)  lenticular  filters,  covering  the wavelength  range  between
160~nm and  600~nm \citep{Poole:2008qy}. The exposure  times vary from
source to source depending on visibility and brightness of the optical
counterpart. Approximate values are 100\,s each in $u, b, v$ and 200, 250,
400\,s in  $uvw1, uvm2, uvw2$, respectively. For  a few sources multiple
consecutive orbits, and thus longer exposures, were obtained.

\begin{figure}
\centering
\includegraphics[width=0.5\textwidth]{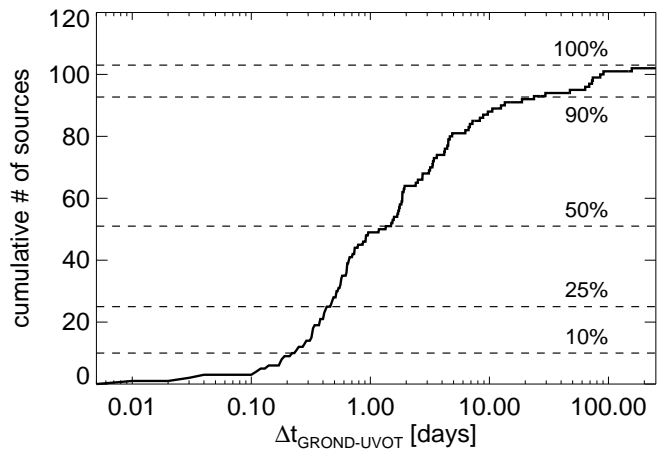}
\caption{Cumulative distribution of the (absolute) time between the
  mid-points of the GROND and {\it Swift}/UVOT observations. The median
offset is 1.5\,d. For longer UVOT observations, which were spread over several
satellite orbits, the mid-time of the $uvw2$-exposure was taken as
 reference.}
\label{fig:offsetdistribution}
\end{figure}

While  UVOT needs  to cycle  through all  filters  in sequence, GROND
observes  in   it's  four  optical   $g^\prime,  r^\prime,  i^\prime,
z^\prime$   and   three   near-IR   $J,   H$,  $K_s$  bands
simultaneously. This capability is particularly important for  studies of
fast-varying sources such as blazars  as it allows the construction of
a reliable spectral energy  distribution (SED) without the need  to correct for
source-intrinsic  variability.  Unfortunately,  scheduling limitations rarely
allowed  simultaneous observations  with both  instruments,  UVOT from
space and GROND from the ground. However, in order to
minimize  the  impact  of  variability  on  the  combined  SED,  GROND
observations were  executed as close as  possible in time  to the {\it
  Swift}  pointings.    Except  in   a  few  cases   where  visibility
or weather constraints  lead to  offsets of  more than  10\,d,  observations were
typically    achieved    within   1-2\,d    of    each   other    (see
Fig.~\ref{fig:offsetdistribution}).  The $z_{\rm  phot}$ uncertainties
associated    with    these    offsets    will   be    discussed    in
Sect.~\ref{sec:varCor}. 

A typical GROND observation had an integration time of 2.4\,min in ($g^\prime,  r^\prime,  i^\prime,
z^\prime$) and 4.0\,min in ($J,   H$, $K_s$).

\section{Data \& Analysis}
\label{sec:data}

\subsection{{\it Swift/UVOT and GROND}}

UVOT  photometry was  carried  out on  pipeline  processed sky  images
downloaded      from     the      {\it     Swift}      data     center~\footnote{
http://www.swift.ac.uk/swift\_portal},  following  the standard  UVOT
procedure \citep{Poole:2008qy}.   Source photometric measurements were
extracted from the UVOT imaging  data using the tool {\sc uvotmaghist}
(v1.1)  with a  circular  source extraction  region  that ranged  from
3\farcs5-5$^{\prime\prime}$  radius to  maximise the  signal-to-noise. In  order  to remain
compatible with the effective area calibrations, which are based on 5"
aperture photometry  \citep{Poole:2008qy}, an aperture  correction was
applied  where necessary.   This correction was at maximum  5--6\,\%
of the flux, depending on the filter. For the  further analysis,  all magnitudes
were converted into the  AB system and are presented in
Tab.~\ref{tab:uvot}. Typically achieved 3-$\sigma$ limiting magnitudes
are $uvw2_{\rm AB}\approx22.4$ and $b_{\rm AB}\approx19.6$.                          

\begin{table*}
\caption{\label{tab:uvot} {\it Swift}/UVOT photometry (first five
  rows, rest in Journal only)}
\begin{tabular}{llccccccc}
\hline\hline
Name  & UT Date$^a$ &\multicolumn{6}{c}{AB Magnitude$^b$} & $\Delta m_{GR\rightarrow UV}$$^c$\\
& & $uvw2$ & $uvm2$ & $uvw1$ & $u$ & $b$ & $v$ & [mag] \\
\hline
CRATES~J0001... & 2010/10/14 10:40 & $18.99\pm0.08$ & $18.74\pm0.10$ & $18.58\pm0.09$ & $18.18\pm0.09$ & $17.88\pm0.10$ & $17.90\pm0.19$ & 0.01 \\
CRATES~J0009... & 2010/12/07 08:55 & $20.16\pm0.15$ & $19.71\pm0.19$ & $19.69\pm0.18$ & $19.24\pm0.19$ & $18.51\pm0.18$ & $18.14\pm0.24$ & -0.24 \\
CRATES~J0021... & 2010/11/18 18:32 & $18.86\pm0.08$ & $18.90\pm0.27$ & $18.57\pm0.09$ & $18.15\pm0.10$ & $17.82\pm0.11$ & $17.93\pm0.21$ & 0.09 \\
1RXS~J002209.2... & 2010/11/18 13:49 & $18.36\pm0.07$ & $17.95\pm0.08$ & $17.91\pm0.08$ & $17.48\pm0.07$ & $17.20\pm0.08$ & $16.97\pm0.12$ & -0.08 \\
RX~J0035.2... & 2011/01/14 01:08 & $18.36\pm0.08$ & $17.95\pm0.08$ & $17.91\pm0.08$ & $17.48\pm0.07$ & $17.20\pm0.08$ & $16.97\pm0.12$ & -0.24 \\
.. & .. & .. & .. & .. & .. & .. & .. & ..\\
\hline
\end{tabular}
\\
$^a$: Approximate start time of $uvw2$ exposure.\\
$^b$: Corrrected for Galactic foreground reddening. Upper limits are 3-$\sigma$.\\
$^c$: Variability-correction factor to be applied to GROND photometry (see text). 
\end{table*}

The GROND  data (Tab.~\ref{tab:grond}) were reduced  and analyzed with
the  standard tools  and methods  described  in \cite{Kruhler:2008aa}.
The  $g^{\prime}, r^{\prime},  i^{\prime}, z^{\prime}$  photometry was
obtained  using  point-spread-function  (PSF) fitting  and  calibrated
against  observations of  fields covered  by the  SDSS Data  Release 8
\citep[][]{Aihara:2011lr}.   Due  to   the  undersampled  PSF  in  the
near-infrared, the $J, H,  K_s$ photometry was measured from apertures
with the sizes corresponding to  the Full-Width at Half Maximum (FWHM)
of   field  stars   and  calibrated   against  selected   2MASS  stars
\citep[][]{Skrutskie:2006wd}.   This resulted in  1$\sigma$ accuracies
of 0.04\,mag ($g^\prime, z^\prime$), 0.03\,mag ($r^\prime, i^\prime$),
0.05\,mag   ($J,  H$),   and  0.07\,mag   ($K_s$)  for   the  absolute
calibration. While  the SDSS calibration  directly provides magnitudes
in   the  AB   system,   the  near-IR   photometry  required   further
transformation from  the 2MASS-native Vega system  into AB.  Typically
achieved    3-$\sigma$   limiting   magnitudes    are
$r^\prime_{\rm AB}\approx23.5$ and $K_{s, \rm
  AB}\approx19.8$.

\begin{table*}
\caption{\label{tab:grond}GROND photometry (first five rows, rest
  in Journal only)}
\begin{tabular}{llccccccc}
\hline\hline
Name  & UT Date$^a$ &\multicolumn{7}{c}{AB Magnitude$^b$} \\
& & $g^\prime$ & $r^\prime$ & $i^\prime$ & $z^\prime$ & $J$ & $H$  & $K_s$\\
\hline
CRATES~J0001... & 2010/10/14 03:33 & $17.80\pm0.05$ & $17.43\pm0.05$ & $17.18\pm0.05$ & $16.86\pm0.05$ & $16.48\pm0.05$ & $16.16\pm0.06$ & $15.76\pm0.08$ \\
CRATES~J0009... & 2010/12/12 00:41 & $18.91\pm0.05$ & $18.49\pm0.05$ & $18.15\pm0.05$ & $17.89\pm0.06$ & $17.34\pm0.06$ & $16.90\pm0.06$ & $16.46\pm0.08$ \\
CRATES~J0021... & 2010/11/23 00:13 & $17.69\pm0.05$ & $17.43\pm0.05$ & $17.20\pm0.05$ & $16.98\pm0.05$ & $16.76\pm0.05$ & $16.51\pm0.05$ & $16.15\pm0.07$ \\
1RXS~J002209.2... & 2010/11/23 00:21 & $17.30\pm0.05$ & $16.92\pm0.05$ & $16.71\pm0.05$ & $16.51\pm0.05$ & $16.22\pm0.05$ & $16.00\pm0.06$ & $15.57\pm0.08$ \\
RX~J0035.2... & 2011/01/14 00:48 & $17.39\pm0.05$ & $17.10\pm0.05$ & $16.96\pm0.05$ & $16.77\pm0.05$ & $16.41\pm0.06$ & $16.17\pm0.05$ & $15.92\pm0.08$ \\
.. & .. & .. & .. & .. & .. & .. & .. & ..\\
\hline
\end{tabular}
\\
$^a$: Exposure start time. \\
$^b$: Corrrected for Galactic foreground reddening. Not corrected for variability. Upper limits are 3-$\sigma$.\\
\end{table*}
 
Correction for Galactic foreground extinction was performed following
the procedure described in \cite{Cardelli:1989lr} with E$_{B-V}$ from
\cite{Schlegel:1998ul}.  For the UVOT bands, the correction factors
presented in \cite{Kataoka:2008fk} were used.  Uncertainties in
  E$_{B-V}$ \citep[10\,\%;][]{Schlegel:1998ul} and in the reddening
  law produce an additional contribution to the photometric error
  budget. As this systematic uncertainty is coupled between the the
  photometric bands for a given position in the sky, its impact on the
  SED fitting is smaller than the contribution to the photometry in
  each individual filter. An exact calculation is very complex. Thus,
  we adopt a conservative 5\,\% of the reddening value in each band
  which is added in quadrature to the photometric uncertainties.

\subsection{Counterpart selection and morphology}

\begin{figure}
   \centering
   \includegraphics[width=0.5\textwidth]{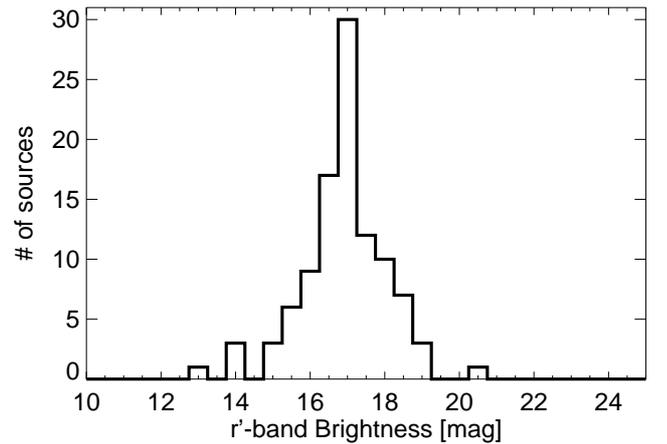}
      \caption{Observed GROND $r^\prime$-band magnitude distribution of the optical counterparts.}
         \label{fig:r-band_distribution}
   \end{figure}

Astrometric solutions  for the GROND optical bands  have been obtained
through comparison with USNO or,  if  available,  with
SDSS measurements, achieving  a  typical rms  of
0\farcs3 in  both coordinates. For  the large majority of  our targets
(91) only a single  optical source was found within 2$^{\prime\prime}$
radius of  the location given in  the 2LAC catalog.  For an additional
twelve  objects  isolated  candidate  counterparts  were  detected  at
offsets of  up to 6$^{\prime\prime}$. In one  case (CGRaBS 1407-4302),
several    blended   sources    made    a   reliable    identification
impossible.   Counterpart  coordinates  and   offsets  for the 103
sources with identification are   given  in
Tab.~\ref{tab:sample}.  Most  of the  identified  optical sources  are
bright   and  point-like   with  typical   magnitudes  in   the  GROND
$r^\prime$-band of 16--18\,mag (Figure~\ref{fig:r-band_distribution}).

The source morphology can be used as a prior for the selection of the
SED template library (see Sect.~\ref{sec:sed}) and redshift range,
i.e., clearly extended sources will have a contribution from the host
galaxy emission to the observed SED and are likely located at low
redshift. A simple morphological distinction in point-like and
extended has been performed on the GROND images and will be used for
discussing the reliability of the $z_{\rm phot}$ estimates.

\subsection{Variability correction}
\label{sec:varCor}

The optical  emission of blazars is known  to vary on a  wide range of
time scales. Brightness variations range from a few tenths of a magnitude
within                 minutes                 to                 days
\citep[e.g.,][]{Racine:1970lr,Miller:1989fk,Carini:1991qy,Urry:1993fj}
to    several    magnitudes    over      weeks    to    years
\citep[e.g.,][]{Ciprini:2003uq,Raiteri:2005kx}.     This   variability
contributes significantly to the  uncertainties when constructing an SED
from non-simultaneous multi-band, multi-instrument observations.

As  GROND  observes  in  all   seven  bands  at  the  same  time,  the
$g^{\prime}$ to $K_s$ photometry
can be considered  as a snapshot of the SED and  is thus unaffected by
variability on time  scales longer than the exposure  of an individual
observation.   This  leaves two  areas  where  a  proper treatment  of
variability has  to be performed,  i) between the exposures in the individual  UVOT filters
and ii) between the GROND and UVOT pointings.

\begin{figure}
   \centering
   \includegraphics[width=0.5\textwidth]{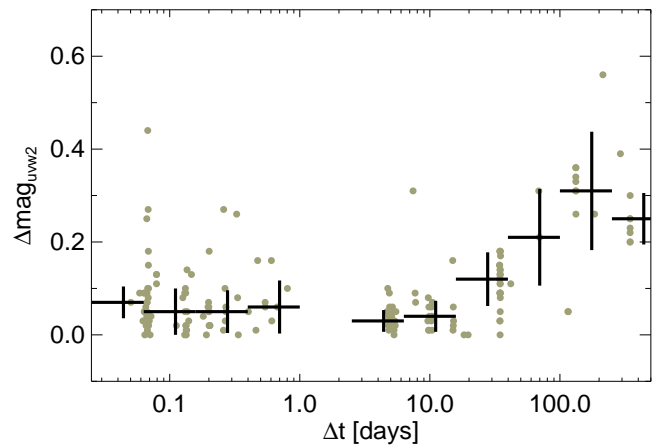}
      \caption{Absolute brightness changes in the $uvw2$ filter as function of time between two UVOT
        pointings for all 25 sources with multiple observations. The black crosses show the median values
      (and median absolute deviation) and the grey dots mark all 174
      pairs of observations. While micro-variability ($\Delta
      m_{uvw2}\approx0.1$\,mag) dominates  on time scales shorter than
      50\,d, significantly larger variability ($\Delta
      m_{uvw2}>0.3$\,mag)  is observed at $\Delta t>50$\,d. }
         \label{fig:variability}
   \end{figure}

UVOT observes in each band  separately and, for our program, typically
cycled through its  filters in a specific order ($uvw1, u  , b , uvw2,
v, uvm2$). A complete sequence in all six bands took $\approx12$\,min,
during which time the target may vary in brightness.  Without a priori
knowledge  of  the  spectral  shape,  this  variability  can  only  be
accounted  for   statistically,  i.e.,  by   including  an  additional
contribution  to the  systematic uncertainties.   For this  purpose we
analyzed  the photometry  for those  blazars for  which  multiple UVOT
exposures in  a single  band were available.  Here, we focused  on the
bluest filter  ($uvw2$) which would  be least affected by  a potential
host galaxy contribution and should thus give a good representation of
the maximum variability.
In   Fig.~\ref{fig:variability}   the   distribution   of
variability as function of time  between two $uvw2$ exposures is shown
for 25  targets with accurate  photometry ($\delta m_{uvw2}<0.1$\,mag)
in   each   image.    While    variability   as   large   as   $\Delta
m_{uvw2}\approx0.4$\,mag  can be seen  for some  sources also  on time
scales of $\Delta t<0.1$\,d, the median variation for $\Delta t<50$\,d
is $\approx0.1$\,mag, comparable to the photometric uncertainty.   

The shortest time sampled between  two $uvw2$ exposures in this sample
is one satellite orbit ($\approx90$\,min) and therefore longer than it
takes to cycle  through the six filters. However,  it is reasonable to
assume that the  median variation on a 12\,min time  scale is not much
larger  than the  one over  96\,min. Thus,  we  conservatively adopted
$\Delta m=0.1$\,mag  as an  additional contribution to  the systematic
uncertainties for all UVOT filters.

\begin{figure}
   \centering
  \includegraphics[width=0.5\textwidth]{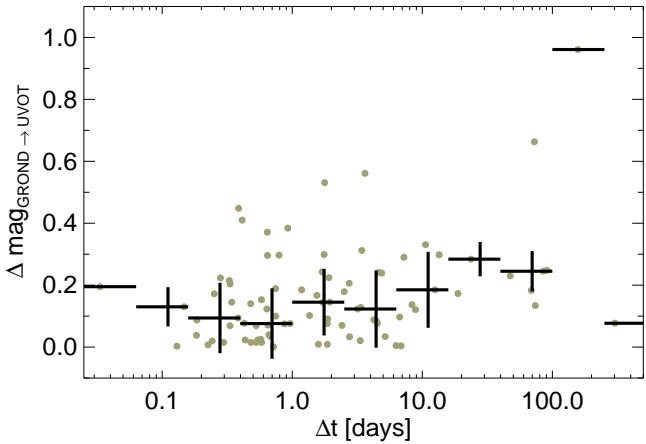}
      \caption{Absolute brightness offset as function of time between the GROND
        and {\it Swift}/UVOT observations. The black crosses show the median values
      (and median absolute deviation) and the grey dots mark all 90
      sources for which the offset was determined (excluding eleven sources
      with large photometric uncertainties, missing b, g$^\prime$,
      or r$^\prime$ photometry, or strong
      host galaxy contribution).}
         \label{fig:corrfac}
   \end{figure}

   In  order to  correct between  GROND and  UVOT we  made use  of the
   spectral    overlap    provided     by    the    two    instruments
   \citep{Kruhler:2011lr}.  Under the  assumption that the SED remains
   unchanged and can be approximated by a power law and thus, that the
   GROND  and  $b,  v$  photometry  follow  the  same  spectral  slope
   \footnote{This is  reasonable for redshift $z<3.5$  and for sources
     where the host galaxy contribution to the emission is negligible.
     See Sect.~\ref{sec:sed} for further discussion.}, color-terms can be
   used  to  derive  the  normalization  offset.   Here  we  used  the
   calibration  with  the largest  spectral  overlap,  namely the  one
   between $b, g^\prime$, and $r^\prime$:

\begin{equation}
b-g^\prime = 0.17(g^\prime-r^\prime)+0.03(g^\prime-r^\prime)^2
\end{equation}

which   is   valid   for    $-1   \le   (g^\prime-r^\prime)   \le   2$
\citep{Kruhler:2011lr}.   These  corrections,  $\Delta  mag_{{\rm  GR}
  \rightarrow {\rm UV}}$ (Tab~\ref{tab:uvot}), were calculated for all
sources with  accurate photometry  in $b, g^{\prime}$,  and $r^\prime$
($\delta  m   <0.2$\,mag).   They   were  computed  for   each  object
individually and applied to the  GROND photometry, shifting it in line
with  UVOT.   Typically  values  are  of the  order  of  $\pm0.1$\,mag
(Fig.~\ref{fig:corrfac}).   In   cases  where  multiple   epochs  were
available,   we   used   the   closest   pair  of   GROND   and   UVOT
observations. The  final SED for  each source corresponds to  the best
possible reconstruction between 160--2200\,nm  at the time of the UVOT
observation.  No  correction  was  performed  for  sources  where  the
photometric uncertainty  ($\delta m \geq0.2$\,mag  in $b, g^{\prime}$,
and/or $r^\prime$)  was larger than the typical  $\Delta mag_{{\rm GR}
  \rightarrow {\rm UV}}$.

   For a  few sources,  the emission from  the underlying  host galaxy
   contributes    significantly    to    the    overall    SED    (see
   Sect.~\ref{sec:sed}).  While  the blazar emission can  still vary, the
   host  contribution  naturally   remains  unchanged,  violating  the
   assumption of a constant SED slope. Thus, no variability correction
   was performed  in those cases where  the UV to  optical SED clearly
   deviates from a power law.

\subsection{SED fitting}
\label{sec:sed}

The photometric redshift computation was performed with the publically
available          {\sc          LePhare}          code          v.2.2
\citep{Arnouts:1999fk,Ilbert:2006qy}.   The  program  uses the  simple
$\chi^2$ fitting method to  differentiate between a theoretical and an
observed  photometric  catalogue.    We  used  three  custom  template
libraries for comparison with our GROND and UVOT photometry. The first
was    composed    of   40    power    law    SEDs    of   the    form
$F_{\lambda}\propto\lambda^{-\beta}$ with $\beta$  ranging from 0 to 2
in steps of 0.05  and describes primarily the central-engine dominated
blazars.  Here, the source-intrinsic extinction was assumed to be zero
and the  luminosity prior was choosen  to cover the  expected range of
absolute                                                     magnitudes
\citep[$-20<M_{g^\prime}<-30$;][]{Veron-Cetty:2010lr}.   Such a simple
model  is a good  approximation for  the UV-near-IR  SEDs of  BL Lacs,
which form the bulk of our sample. Also FSRQs, if their emission lines
are not dominant, can be modeled  with a power law to first order. The
second library  contains templates  of normal (inactive)  galaxies and
galaxy/AGN hybrids \citep{Salvato:2009aa, Salvato:2011lr} and has been
implemented in order  to model host galaxy dominated  objects.  If the
host contribution is significant,  the 4000\,\AA\ break can emerge and
potentially be misinterpreted as  a Lyman-limit.  While for the galaxy
models   intrinsic  extinction  was   again  neglected,   a  different
luminosity prior of $-8<M_{g^\prime}<-30$  was chosen to allow for the
expected  fainter host magnitudes.   The third  library consists  of a
wide           range           of          stellar           templates
\citep[][]{Pickles:1998xy,Bohlin:1995qy,Chabrier:2000uq},  included to
test  against potential false  associations.  For  each source  in our
sample, all three libraries were fit independently.

The most prominent spectral feature used to measure $z_{\rm phot}$ for
blazars is caused by the absorption through neutral hydrogen along the
line  of  sight.  At  $z\approx0.8$  the  Lyman-limit  moves into  the
wavelength covered by the UVOT and  starts to suppress the flux in the
$uvw2$ filter.  While  this is the theoretical lower  limit on $z_{\rm
  phot}$  for  the combined  GROND  and  UVOT coverage,  uncertainties
associated with the photometry and  modeling will shift it to a higher
redshift.  Usually one would  use a  spectroscopic training  sample to
evaluate the redshift range and  accuracy which is accessible with our
SEDs.   As the  majority of  the sources  with known  redshift  in our
sample are  located at $z<1$, they  alone do not suffice  to provide a
statistically meaningful  test of the  method.  Instead we  assess the
reliability in our photometric redshift computation with a Monte-Carlo
approach.

Here,  we simulated 27,000  test power-law  SEDs with  spectral slopes
ranging from  0.5 to 2, redshifts from  0 to 4, and  at three apparent
magnitudes,   $r^{\prime}=17,18,19$\,mag.   For   each   source,   the
photometry  in the  individual bands  was scattered  around  the model
magnitudes  with a  brightness-dependent statistical  contribution and
with a  component representing  the systematic uncertainties  from the
photometric calibration and variability correction. The resulting SEDs
were fed  back into  {\sc LePhare} and  the output $z_{\rm  phot}$ was
compared to the input value.

\begin{figure}
   \centering
   \includegraphics[width=0.5\textwidth]{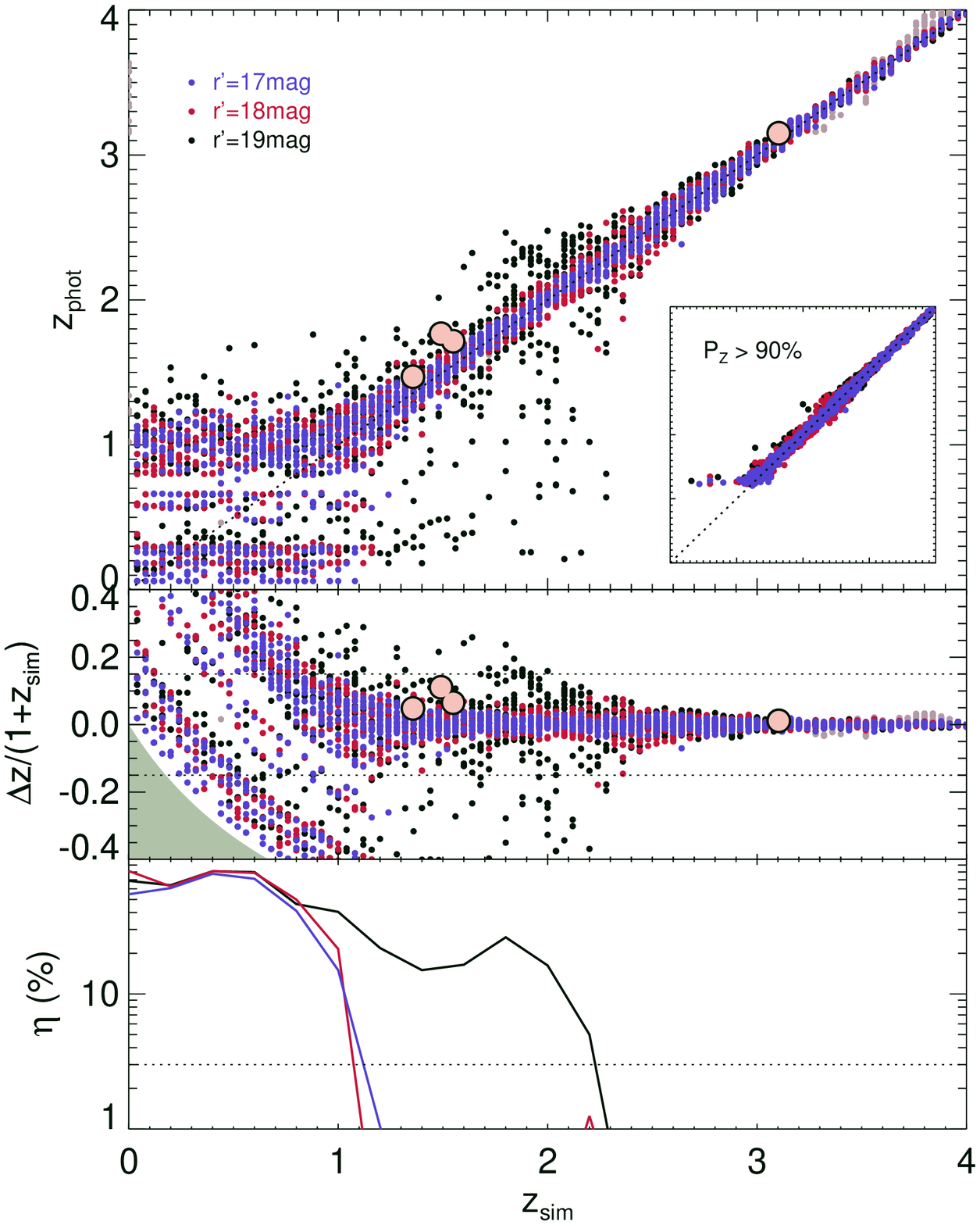}
      \caption{Recovered best-fit photometric vs input redshift (top),
        accuracy (middle), and outlier fraction (bottom) for 27,000 simulated sources with spectral
        slope of $0.05\le\beta\le2$ and optical brightness of
        $r^\prime=17,18,19$\,mag.  The pink circles indicate the
        location of three blazars in our sample with known redshift, \object{PKS~0332-403} \citep[$z_{\rm
   spec}=1.351$,][]{Bergeron:2011lr}, \object{OM~235} ($z_{\rm spec}=1.549$),
 and \object{CRATES~J1312-2156} \citep[$z_{\rm
   spec}=1.491$,][]{Ackermann:2011lr}, as well as that of  \object{PKS~0537-286} \citep[$z_{\rm
   spec}=3.104$,][]{Wright:1978lr}, a LAT-non-detected blazar
 with available GROND and UVOT photometry.
 The green area in the middle
 panel corresponds to the forbidden parameter space where 
 $z_{\rm phot}>0$. Systematics are visible at $z_{\rm phot}<0.8$. The
 large majority of the  fits have $\chi^2<30$ (for typically 10 d.o.f.)
 with the exceptions shown in light grey. The inset in the top panel displays the results if only solutions with P$_{\rm z}>90$\,\% are selected.
}
         \label{fig:photozSimPL}
   \end{figure}

As  can  be  seen  in  Fig.~\ref{fig:photozSimPL},  there  is  a  good
correspondence  between  the  input  and the  recovered  redshift  for
$z_{\rm    sim}>1.2$.    In     particular    for    bright    sources
($r^\prime<18$\,mag)  the number  of outliers  $\eta$, defined  as the
number of sources with $|\Delta z(1+z_{\rm sim})|>0.15$, drops steeply
above this  redshift. For fainter  sources ($r^\prime\approx19$\,mag),
the  increasingly less  constrained UV  photometry causes  the outlier
fraction   to   remain   high   ($\approx20$\,\%)   out   to   $z_{\rm
  sim}\approx2$.  
A more quantitative selection can  be made when using P$_{\rm z}$, the
integral of  the probability distribution  function $\int f(z)  dz$ at
$z_{\rm phot} \pm0.1(1+z_{\rm phot})$, which describes the probability
that the  redshift of a  source is within  $0.1(1+z)$ of the  best fit
value. When choosing a cut  at P$_{\rm z}>90$\,\%, the majority of the
outliers at  $z_{\rm phot}>1.2$, as  well as nearly all  the solutions
with         $z_{\rm        phot}<1.2$         disappear        (inset
Fig.~\ref{fig:photozSimPL}). This  confims that for  power-law sources
with $z<1.2$ the  available photometry can only place  an upper limit,
and suggests  that P$_{\rm  z}>90$\,\% can be  used as a  criterion to
identify  reliable photometric  redshift solutions.  Selecting a
  higher $P_{\rm  z}$ threshold, e.g., 99\,\%, can  further reduce the
  outlier  fraction  but  will  also  shift  the  lower  constrainable
  redshift  bound to  higher values.  In order  to reach  such  a very
  tightly peaked redshift  probability distribution function more than
  one photometric band has to be affected by the Lyman-limit, limiting
the redshift range to $z\gtrsim1.5$.

\section{Results}
\label{sec:results}

The fit results for all 103 sources with identified counterparts are presented in
Tab.~\ref{tab:result}. Here we give the photometric redshifts (and
their 90\,\% confidence errors), the P$_{\rm z}$ values, and the best fit models for
the power law and galaxy templates. No source requires  a stellar template.

\begin{table*}[tb]
\caption{\label{tab:result}Excerpt of the results table. Only entries
  for those sources with reliable redshifts of $z_{\rm phot}>1.2$ are
  shown. (rest in Journal only)}
\begin{tabular}{lcc|cccc|cccl}
\hline\hline
Name  & $z_{\rm phot, best}$$^a$ & $z_{\rm spec, img}$$^b$  &\multicolumn{4}{c|}{power law} & \multicolumn{4}{c}{galaxy} \\
& & & $z_{\rm phot}$ $^c$ & $\chi^2$ & P$_{\rm z}$ $^d$ & $\beta^e$ &
$z_{\rm phot}$ $^c$& $\chi^2$ & P$_{\rm z}$  $^c$ & model \\
\hline
RX~J0035.2+1515 & 1.28$_{-0.17}^{+0.14}$ & -- & 1.28$_{-0.17}^{+0.14}$ & 2.6 & 90.1 & 0.90 & 0.79$_{-0.13}^{+0.14}$ & 8.7 & 75.5 & I22491\_60\_TQSO1\_40\\
PKS~0047+023 & 1.44$_{-0.19}^{+0.16}$ & -- & 1.44$_{-0.19}^{+0.16}$ & 10.8 & 94.9 & 1.60 & 0.41$_{-0.08}^{+0.06}$ & 13.6 & 80.5 & pl\_QSO\_DR2\_029\_t0\\
PKS~0055-328 & 1.37$_{-0.17}^{+0.14}$ & -- & 1.37$_{-0.17}^{+0.14}$ & 17.7 & 95.2 & 1.25 & 0.47$_{-0.47}^{+0.23}$ & 11.2 & 38.4 & I22491\_50\_TQSO1\_50\\
PKS~0332-403 & 1.47$_{-0.12}^{+0.11}$ & 1.426 & 1.47$_{-0.12}^{+0.11}$ & 6.0 & 99.8 & 1.35 & 1.11$_{-0.33}^{+0.10}$ & 13.0 & 62.3 & I22491\_60\_TQSO1\_40\\
CRATES~J0402-2615 & 1.92$_{-0.09}^{+0.12}$& -- & 1.92$_{-0.09}^{+0.12}$ & 8.6 & 100.0 & 1.25 & 1.28$_{-0.23}^{+0.22}$ & 9.5 & 48.9 & I22491\_60\_TQSO1\_40\\
SUMSS J053748~571828 & 1.55$_{-0.13}^{+0.09}$ & -- & 1.55$_{-0.13}^{+0.09}$ & 20.5 & 99.9 & 0.80 & 1.58$_{-0.07}^{+0.08}$ & 21.5 & 99.6 & pl\_I22491\_30\_TQSO1\_70\\
PKS~0600-749 & 1.54$_{-0.19}^{+0.14}$ & -- & 1.54$_{-0.19}^{+0.14}$ & 6.9 & 98.1 & 1.20 & 0.46$_{-0.05}^{+0.24}$ & 7.5 & 58.1 & I22491\_60\_TQSO1\_40\\
CRATES~J0630-2406 & 1.60$_{-0.05}^{+0.10}$ & -- & 1.60$_{-0.05}^{+0.10}$ & 9.3 & 100.0 & 0.85 & 1.19$_{-0.12}^{+0.50}$ & 24.6 & 83.7 & I22491\_60\_TQSO1\_40\\
OM~235 & 1.72$_{-0.13}^{+0.13}$  & 1.549 & 1.72$_{-0.13}^{+0.13}$ & 18.2 & 99.9 & 1.10 & 1.54$_{-0.09}^{+0.10}$ & 11.9 & 94.9 & pl\_I22491\_30\_TQSO1\_70\\
CRATES~J1312-2156 & 1.77$_{-0.11}^{+0.09}$ & 1.491 & 1.77$_{-0.11}^{+0.09}$ & 7.1 & 100.0 & 0.95 & 1.60$_{-0.09}^{+0.14}$ & 18.4 & 100.0 & I22491\_60\_TQSO1\_40\\
CLASS~J2352+1749 & 1.45$_{-0.18}^{+0.21}$ & -- & 1.45$_{-0.18}^{+0.21}$ & 7.5 & 91.8 & 1.15 & 0.55$_{-0.14}^{+0.16}$ & 7.1 & 64.2 & pl\_I22491\_20\_TQSO1\_80\\
\hline
\end{tabular}
\\
$^a$: Best photometric redshift, see text.\\
$^b$: Spectroscopic or imaging redshift (if known) from \cite{Ackermann:2011lr}.\\
$^c$: Photometric redshift with 90\,\% confidence uncertainties. \\
$^d$: Redshift probability density at $z_{\rm phot} \pm0.1(1+z_{\rm phot})$. \\
$^e$: Spectral slope for power law model of the form  $\lambda^{-\beta}$.\\
$^f$: Starburst/QSO hybrid templates with varying contributions of the
two components. Template names reflect the relative contribution of
the starburst (I22491\_NN) and AGN (TQSO1\_NN) models to the
hybrid. ``pl'' marks an additional power-law component at short
wavelengths. See \cite{Salvato:2011lr} for
details on the templates.\\
\end{table*}
 
As discussed in  the previous section, P$_{\rm z}>90$\,\%  can be used
as  a   good  reliability  criterion  for   the  photometric  redshift
solution. Applying this  cut to the power law model  fits results in a
sample of  15 sources with $\chi^2<30$. Next, we also require that a
source can not be fit well with a galaxy/AGN-hybrid template at low
redshift ($z<1.2$). This step is required in order to identify those
sources for which a degeneracy in the photometric redshift solution
does not allow a distinction between a low-redshift galaxy/AGN-hybrid
and a blazar at higher redshift. In analogy with our selection for a
good redshift constraint (P$_{\rm z}>90$\,\%), we consider P$_{\rm z,
  gal}<90$\,\% as a reliable criterion that no adequate low-redshift
galaxy/AGN-hybrid solution was found. This removes further seven
sources from the sample and leaves eight  candidates, all  unresolved in  the GROND
images and with $z_{\rm phot, pl}>1.2$.

\begin{figure}[h]
   \centering
  \includegraphics[width=0.5\textwidth]{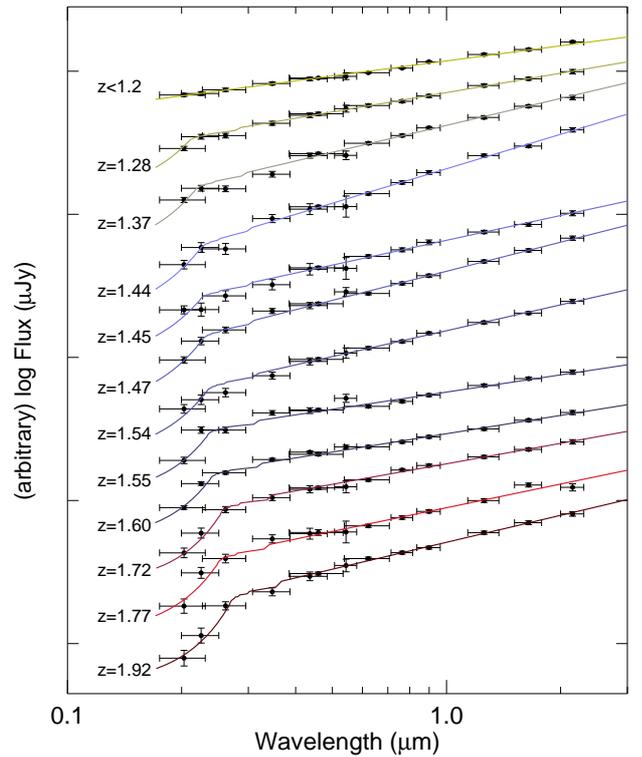}
      \caption{SEDs for the eleven sources for which photometric redshifts larger than 1.2 were estimated (see Tab.~\ref{tab:result}) and one example for $z_{\rm phot}<1.2$. From bottom to top: \object{CRATES~J0402-2615}, \object{CRATES~J1312-2156}, \object{OM~235}, \object{CRATES~J0630-2406}, \object{SUMSS~J053748-571828}, \object{PKS~0600-749}, \object{PKS~0332-403}, \object{CLASS~J2352+1749}, \object{PKS~0047+023}, \object{PKS~0055-328}, \object{RX~J0035.2+1515}, and \object{BZB~J0543-5532}.}
         \label{fig:SEDs}
   \end{figure}

For  three additional  sources the  power  law and  galaxy models  fit
similarly well  (P$_{\rm z,  pl}\sim{\rm P}_{\rm z,  gal}>90$\,\%) and
give   comparable  high-redshift   solutions,   indicating  that   the
Lyman-limit was fit by power law and galaxy templates alike. Including
those,  we end  up  with a  sample  of eleven  sources with reliable
photometric redshift computation, all , with  $z_{\rm phot}$  ranging   from  $z_{\rm  phot}\approx1.28$   to  $z_{\rm
  phot}\approx1.92$.  This  encompasses  eight sources  without  previous
redshift measurements  and three blazars with known  redshift. For the
latter, the photometric redshifts are within 3-$\sigma$ of the known
value  (see Fig.~\ref{fig:photozSimPL}), indicating an accuracy of $\Delta
z  / (1+z_{\rm  spec})<0.12$ for our $z_{\rm phot}$ method
described in Sect.~\ref{sec:sed}.  The fit results and SEDs for
the  eleven   sources  are  presented   in  Tab.~\ref{tab:result}  and
Fig.~\ref{fig:SEDs}, respectively.

   For 21  sources, the  galaxy library provides  the better  fit with
   P$_{\rm z,  gal}>90$\,\%, P$_{\rm z,  pl}<90$\,\%, and $\chi^2_{\rm
     gal}<30$. All of these are best described by starburst/QSO hybrid
   templates at $z<1.3$ and none require an elliptical galaxy, i.e., a
   strong  4000\,\AA\ break. The  most significant  difference between
   the starburst/QSO hybrid templates and the power-law models is that
   the   former  exhibit  strong   emission  lines   originating  from
   star-formation   \citep[see   ][    for   more   details   on   the
   templates]{Salvato:2009aa}.   These  lines  can  be  fit  to  small
   deviations in the photometry from  a power-law and thus provide low
   $\chi^2$  values  and  apparently well-constrained  $z_{\rm  phot}$
   solutions.   However,  as  described  in  Sect.~\ref{sec:varCor},  the
   variability   correction   is   not   reliable  when   applied   to
   non-power-law sources,  as the color  terms, and thus  the correction
   factors,  will  be  erroneous.   Also,  18 of  these  sources  have
   previously  been classified  as  BL~Lac objects  and are  therefore
   unlikely to  show strong emission  lines in their  optical spectra.
   Given these considerations, we  do not regard their galaxy-template
   photometric redshifts to be reliable.

   There  are 79  sources which  are well-fit  by  power-law templates
   ($\chi^2<30$)  but have  a low  P$_{\rm  z}$. In  other words,  the
   spectral slope is well  defined but the photometric redshift cannot
   be unambiguously determined.  As expected, the best solutions
   are predominantly  at low redshift ($z<1.2$)  with typical redshift
   probability distributions that are flat down to $z=0$. For those we
   give  the 90\,\% upper  limit of  the photometric  redshift..  This
   sub-sample  also includes  eleven sources  with a  previously known
   redshift from  spectroscopy or imaging,  all in agreement  with our
   limits.

   We also provide  the 90\,\% upper limit for  four sources for which
   the power-law and galaxy  libraries give similarly good constraints
   (P$_{\rm   z,pl}\approx$P$_{\rm  z,gal}>90$\,\%)   but   with  very
   different redshift solutions. This degeneracy can occur, e.g., when
   a break in  the SED is similarly  well fit by the Lyman  limit of a
   power-law model and with the 4000\,\AA\ break of a galaxy template.
   No  $z_{\rm phot,  best}$ is  given for  ten sources  for  which no
   satisfactory fit ($\chi^2>30$) was obtained.

\section{Discussion \& Conclusion}
\label{sec:discussion}

\begin{figure}
   \centering
   \includegraphics[width=0.5\textwidth]{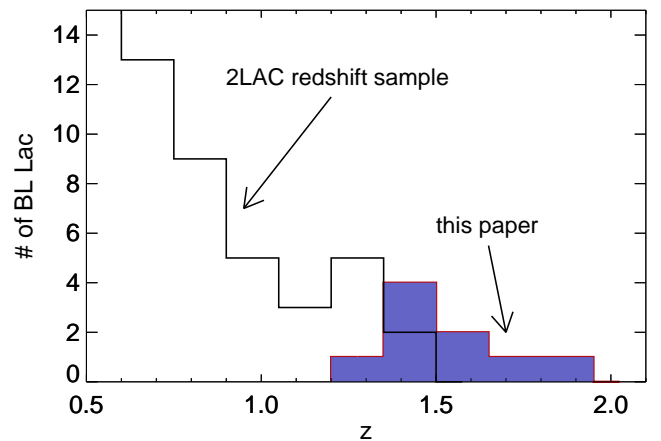}
      \caption{Redshift distribution for the nine BL Lac objects with
        reliable photometric redshifts (blue filled histogram)
        together with the distribution for the BL~Lac in the 2LAC catalog
        \citep[empty histogram;][]{Ackermann:2011lr}. Sources with
        $z<0.5$ have been omitted for clarity.}
         \label{fig:zDis}
   \end{figure}

In this paper  we presented redshift constraints for  103 blazars from
the  2LAC catalog using  UV-to-near-IR multi-band  photometry obtained
quasi-simultaneously with {\it Swift}/UVOT and GROND.  We provided the
first reliable  redshift measurements for eight sources  and new upper
limits  for  an additional  66  targets.  Of  the eight  sources  with
reliable  redshift, seven are  located at  $z_{\rm phot}>1.3$.  Six of
those belong  to the  BL Lac population.   For comparison, out  of the
total 395 BL  Lac in the 2LAC sample only  two sources have previously
been  known to  lie at  $z>1.3$ (Fig.~\ref{fig:zDis}).   Redshifts for
these  two,   \object{PKS~0332-403}  and  \object{CRATES~J1312-2156},   have  also  been
confirmed with our photometric observations.

   The six new BL Lac  redshifts at $z>1.3$ represent a dramatic (from
   two to eight)  increase of the confirmed high-z  {\it Fermi} sample
   and thus demonstrate the  opportunity that the SED template fitting
   technique  holds for  obtaining  photometric redshifts  for BL  Lac
   sources.    For  our  sample,   this  was   possible  due   to  the
   densely-covered wide spectral  range (160--2200\,nm), necessary for
   a  reliable  constraint  of   the  spectral  slope,  the  excellent
   ultra-violet   coverage  to  measure   the  Lyman-limit,   and  the
   quasi-simultaneity  of the  observations, important  for minimizing
   the impact of source-intrinsic  variability.  The method applied in
   this  work  overcomes  significant  challenges  inherent  in  other
   redshift techniques, namely the  simplicity of the optical emission
   of BL Lacs manifested  as the power-law-shaped synchrotron spectral
   component.   This complicates spectroscopic  redshift measurements,
   which are  in most cases  limited to the optical  wavelength regime
   and  thus  insensitive to  the  Lyman-limit. Instead,  spectroscopy
   relies  on the detection  of very  faint emission  features, mainly
   from the  underlying host galaxy.  It therefore requires  very high
   signal-to-noise, often at  the cost of long exposure  times. On the
   other hand,  the photometric  redshift method for  BL Lac  does not
   require significant  telescope time,  in particular when  data from
   efficient multi-channel instruments  like GROND are available.  Its
   weakness, however, is that the  method relies on the detection of a
   particular spectral  feature, which,  albeit strong, is  located in
   the  rest-frame far-ultra-violet.  For $z\lesssim3$,  this  is only
   accessible with space-based  ultra-violet telescopes, and even then
   usually limited to redshifts above $z\approx1.2$, as shown above.

   The sensitivity of this method to high-redshift sources also allows
   the placement of upper limits  for those SEDs that do not show
   the  imprint of  the Lyman-limit.  Only for  three targets  is this
   boundary above $z=2$. In one case, \object{CRATES~J0250+1708} ($z<3.1$), the
   optical counterpart  is too faint  ($r^\prime\approx20.7$\,mag) and
   only  upper limits in  all six  UVOT bands  could be  obtained. The
   counterpart of \object{CRATES~J0705-4847} is  only detected in the GROND $J$
   ($\approx19.7$\,mag$_{\rm  AB}$)  and $H$  bands  and  its SED  and
   redshift  are   thus  poorly  constrained.  Finally,   the  SED  of
   \object{ATG20~J0124-0625}   ($z<2.46$)   shows   a  significant   break   at
   $z\sim1.9\pm0.5$. However, as the redshift probability distribution
   is broad, and the $\chi^2$ of  the power law fit comparable to that
   of a $z\approx0.4$ solution  for a galaxy template, the photometric
   redshift is considered to be unreliable. Except for those three, no
   other source in  our sample has a best-fit  photometric redshift at
   $z>2$.  Thus,  \object{CRATES~J0402-2615} can  now  be  considered the  most
   distant known BL Lac with a measured redshift of $z\approx1.92$.

   Fig.~\ref{fig:zDis}  indicates  that  our  high-redshift findings  are  the
   natural extension  of the  existing 2LAC redshift  sample. However,
   due to the incompletenesses of both samples, we refrain from drawing
   any   quantitative   conclusions   at   this  stage.  We note,
   however, the our result is general agreement with the theoretical
   predictions from \cite{Giommi:2011lr}.  A  more detailed physical
   interpretation  of the  establishment of  an increased  fraction of
   high-z BL Lacs in the 2LAC sample will be reported separately.

\begin{acknowledgements}
We thank the referee for the valuable comments. Part of the funding for GROND (both hardware
as well as personnel) was generously granted from the Leibniz-Prize to
Prof.   G.   Hasinger  (DFG  grant  HA~1850/28-1).  MS acknowledges support by the German Deutsche
Forschungsgemeinschaft, DFG Leibniz Prize (FKZ HA 1850/28-1).
TK acknowledges support  by the DFG cluster of
excellence Origin  and Structure  of the Universe, by the European Commission under the Marie
Curie Intra-European Fellowship Programme), as well as the DARK: The
Dark Cosmology Centre, funded by the Danish National Research Foundation.
FOE acknowledges funding of his Ph.D. through the Deutscher
Akademischer Austausch-Dienst (DAAD). SK, DAK and ANG acknowledge
support by DFG grant Kl 766/16-1. ARossi acknowledges support from the
BLANCEFLOR Boncompagni-Ludovisi, n\'e Bildt foundation. MN
acknowledges support by DFG grant SA 2001/2-1. PS acknowledges support
by DFG grant SA 2001/1-1.  ACU, ANG, DAK and ARossi are grateful for
travel funding support through MPE. We also acknowledge the use of  the TOPCAT tool (Taylor 2005).

\end{acknowledgements}

\bibliographystyle{aa}

\end{document}